\documentclass[aps,prb,twocolumn]{revtex4}
\usepackage{graphics,epsfig,amsfonts,amssymb,amsmath}
\begin{document}
\title{
Energy scales in the Raman spectrum of electron and hole doped cuprates within
competing scenarios
 }
\author{B. Valenzuela$^{1,2}$}\author{E. Bascones$^1$}
\affiliation{$^1$Instituto de Ciencia de Materiales de Madrid, CSIC.
Cantoblanco. E-28049 Madrid, Spain} \affiliation{$^2$Departamento de
F\'{\i}sica de la Materia Condensada, Universidad Aut\'onoma de
Madrid. Cantoblanco. E-28049 Madrid, Spain}
\begin{abstract}
Recent experiments in underdoped hole-doped cuprates have shown the
presence of two energy scales in the Raman spectrum in the
superconducting state. This feature has a natural explanation in
some models in which pseudogap and superconductivity compete. In
electron-doped cuprates antiferromagnetic correlations are believed
to survive in the superconducting state, and to produce a pseudogap
above the critical temperature. Contrary to hole-doped systems, in
electron-doped compounds  only one energy scale appear since
the pair breaking Raman intensity peaks in both B$_{1g}$ (antinodal)
and B$_{2g}$ (nodal) channels at a frequency of a few meV, typical
of the superconducting order parameter. In this paper we analyze the
different effect in the Raman spectrum of the competition between
pseudogap and superconductivity in electron and hole-doped cuprates.
The difference in energy scales in both systems is explained in
terms of the different truncation of the Fermi surface induced by
the pseudogap. For electron-doped cuprates we also analyze the
spectrum with antiferromagnetism and a non-monotonic superconducting
order parameter.

\end{abstract}
\email{belenv@icmm.csic.es,leni@icmm.csic.es}
\date{\today}
\maketitle
\section{Introduction}
The pseudogap (PG) and the asymmetry between electron and hole-doped
cuprates are key issues in high-temperature superconductivity. In
angle-resolved photoemission (ARPES) in hole-doped compounds the PG
manifests in reduced intensity in the antinodal region (close to
$(0,\pi)$) and a Fermi arc in the nodal one, around $(\pi/2,\pi/2)$,
instead of a complete Fermi surface\cite{fermiarc}(FS). Recent
ARPES\cite{twoscalesarpes} and Raman\cite{leTacon06}
 experiments
suggest that the nodal-antinodal dichotomy remains in the
superconducting (SC) state in the form of two energy scales.

Inelastic Raman scattering\cite{ramanreview} permits to
differentiate the zero-momentum charge excitations of nodal
($\chi_{B_{2g}}$) and antinodal regions ($\chi_{B_{1g}}$). In the SC
state, pair breaking peaks appear in the Raman spectrum. In
hole-doped cuprates the $B_{2g}$ peak frequency shows a
non-monotonic dependence on doping $x$, similar to the one of the
critical temperature $T_c$. On the contrary, the $B_{1g}$ intensity
strongly decreases with underdoping with a peak frequency which
seems to evolve from the SC to the PG
scale\cite{leTacon06,nosotras07}. This behavior has been interpreted
as a signature of the competition between SC and
PG\cite{nosotras07,zeyherygreco02}. For alternative descriptions
see\cite{leTacon06,vertex}. Very recent
measurements\cite{sacuto08b1g} show that in underdoped compounds the
position of the $B_{1g}$ peak barely changes with temperature and
that intensity at this frequency remains above $T_c$, opposite to
what happens in overdoped samples. It also appears at the same
frequency in impurity-substituted samples
  with different $T_c$ but the same nominal doping\cite{sacutoimpurezas}.
On the contrary, the nodal
$B_{2g}$ peak displays a significant temperature dependence below
$T_c$ in both the underdoped and the overdoped
regimes\cite{sacuto08b2g}.

In electron-doped cuprates the PG suppresses the ARPES intensity at
the hot spots\cite{armitage01}, where the FS cuts the
antiferromagnetic zone border (AFZB). This suppression remains in
the SC state. Band folding across the AFZB\cite{park07} and a gap of
$\sim 100$ meV at the hot spots have been
observed\cite{park07,matsui05af,aclaraciongap}. This gap remains in
the superconducting state. These features are well reproduced by  a
spin density wave (SDW) state and its coexistence with
superconductivity below $T_c$\cite{mark1,nazarioysantiago}. The SDW
model\cite{Schrieffer} is based on an itinerant electron approach to
the antiferromagnetism. The SDW truncates the FS into electron and
hole pockets at the antinodal and nodal regions. The pockets picture
and the AF origin\cite{tremblay} of the PG agrees with the doping
evolution of the Hall coefficient\cite{hall}, the elastic peaks at
${\bf Q}=(\pi,\pi)$ observed in neutron scattering\cite{neutrones},
magnetotransport\cite{magnetotransport} and optical
conductivity\cite{bontemps} experiments.

Contrary to hole-doped materials, in electron-doped compounds the
pair breaking Raman intensity peaks in both B$_{1g}$ (antinodal) and
B$_{2g}$ (nodal) channels at a frequency of a few meV, typical of
the SC order parameter\cite{blumberg02,qazilbash05}. The peak
frequency has a non-monotonic dependence on doping, similar to the
one of $T_c$. For some dopings $B_{2g}$ peaks at a frequency larger
than $B_{1g}$, what has been interpreted in terms of a non-monotonic
d-wave  SC gap\cite{blumberg02,eremin08,guinea04} with maximum value at
the hot spots\cite{krotov06}. A non-monotonic leading edge shift
below $T_c$ has been observed\cite{matsui05nonm,mark2} in samples which
show a gap of ~100 meV at the hot spots above and below the critical
temperature. Yuan and Yuan\cite{yuanyuan} and Liu {\it et
al}\cite{chinos} respectively proposed that the non-monotonicity of
the ARPES gap and the relative position of the $B_{1g}$ and $B_{2g}$
peaks were a consequence of the coexistence of antiferromagnetism
and d-wave superconductivity. In electron-doped cuprates the SDW and
SC scales can be decoupled\cite{chinosdelasnarices}.  The different
values of SC scale\cite{matsui05nonm,tunel}$\sim 4$ mev with respect
to the pseudogap $\sim 100 meV$ suggest that the non-monotonicity is
not associated with the opening of an antiferromagnetic
gap\cite{aclaracion}. In the presence of a PG as the one seen in
ARPES\cite{matsui05af,park07} the Raman spectrum including both SDW
and a non-monotonic d-wave gap should be studied.

In this article we analyze the different Raman spectrum of electron
and hole doped cuprates within competing scenarios. In
electron-doped cuprates the PG is modeled by a SDW and it is assumed
to remain present in the SC
state\cite{nazarioysantiago,mark1,yuanyuan}. For hole-doped systems
we use the Yang-Rice-Zhang (YRZ) model\cite{YRZ} which reproduces
well\cite{nosotras07} the doping dependencies of $B_{1g}$ and
$B_{2g}$ peak frequencies and intensities. As it is also
experimentally observed, in the hole-doped superconductors the PG
scale couples with the SC one and affects the pair breaking Raman
spectrum, while PG and SC scales are decoupled in electron-doped
compounds. We show that such a difference is not a consequence of
the different model used but of the different region of the Fermi
surface which is truncated by the pseudogap. We also calculate the
Raman spectrum corresponding to a non-monotonic SC d-wave gap, in
the presence of a SDW  to make a closer comparison with experiments.

\section{The model}
{\it Electron-doped cuprates.}
We start from the Green's function in presence of antiferromagnetism
and superconductivity which couples the operators $(c\dag_{{\bf
k},\uparrow},c_{-{\bf k},\downarrow},c\dag_{{\bf
k+Q},\uparrow},c_{{-\bf k-Q},\downarrow})$

\begin{displaymath}
\mathbf{G^{-1}(\omega,{\bf k})}=\left(\begin{array}{c c c
c}\omega-\xi_{\bf k} & -\Delta_{S,\bf k} & -\Delta_{AF} &
0 \\
-\Delta_{S,\bf k}& \omega+\xi_{\bf k} & 0 & -\Delta_{AF}
\\
 -\Delta_{AF}  & 0 & \omega-\xi_{\bf k+Q} &
-\Delta_{S,{\bf k+Q}} \\
0 & -\Delta_{AF} & -\Delta_{S,\bf k+Q} & \omega+\xi_{\bf k+Q}
\end{array}\right)
 \label{eq:green}.
\end{displaymath}
Such a Green's function can be derived from the Hubbard or t-J model
at mean field level\cite{mark1,nazarioysantiago,yuanyuan}. We assume
a doping dependent\cite{mark1} isotropic AF gap $\Delta_{AF}$ and a
d-wave SC order parameter $\Delta_{S,\bf k}=(\Delta_{S}/2 )(\cos k_x
-\cos k_y)$ except otherwise indicated. The band dispersion is
$\xi_{\bf k}= -2t_0 (\cos k_x+\cos k_y)-4t_1 \cos k_x\cos k_y -2t_2
(\cos 2k_x +\cos 2k_y)-\mu$.

Next we consider the Raman response. We use the symmetry of the
point group transformations of the crystal to classify the
scattering amplitude\cite{devereauxkampfprb99}. Since we are mostly
concerned in the anisotropic properties of the system we will
calculate the $B_{1g}$ and the $B_{2g}$ channels and not the
$A_{1g}$ channel. This will be also valid for the hole doped case.
In the bubble approximation the Raman response is:
\begin{eqnarray}
\chi^\nu(\Omega)&=&\frac{1}{N}\sum_{\bf k} (\gamma^\nu _{\bf k})^2
\{ \Pi_{11,11}({\bf k},\Omega)-\Pi_{12,21}({\bf k},\Omega)+
{}\nonumber\\{}&& (-1)^\eta(\Pi_{13,31}({\bf
k},\Omega)-\Pi_{14,41}({\bf k},\Omega)) \}, \label{formularaman}
\end{eqnarray}
\begin{equation}
\Pi_{ij,kl}({\bf k},i\Omega)=T\sum_n G_{ij}(i\omega_n+i\Omega,{\bf
k})G_{kl}(i\omega_n,{\bf k})).
\end{equation}
Here $\gamma_{B_{1g}}\propto (\cos k_x-\cos k_y)$ and
$\gamma_{B_{2g}}\propto (\sin k_x\sin k_y)$ are the Raman vertices,
\footnote{In principle, since the dispersion relation has terms 
of the form $\cos 2k_x +\cos 2k_y$ the $B_{1g}$ Raman vertex could 
also have a contribution of the higher harmonic $\cos 2k_x -\cos 2k_y $.
We have checked that this contribution does not change qualitatively the 
results affecting only to the intensity of the Raman response.}
$\eta=1,2$ respectively for $B_{1g}$ and $B_{2g}$ and the Green's
functions are written
in the extended Brillouin zone. Equivalent expressions
apply\cite{zeyherygreco02} for coexisting d-density wave and SC. At
zero temperature and zero scattering rate:
\begin{equation}
\chi^{B_{2g}}(\Omega)= \frac{1}{4}\sum_{\bf k,\tau=\pm} \left (\gamma^{B_{2g}}_{\bf
k}\right )^2 \frac{\Delta_{S,{\bf k}}^2}{\left (E_{\bf
k}^\tau \right)^2}\left (1+\frac{\tau \xi^-_{\bf k}}{E_{\bf k}}\right ) \delta
   \left (\Omega-2E^\tau_{\bf k}\right),
\label{ramancoexb2g}
\end{equation}
\begin{eqnarray}
\nonumber \chi^{B_{1g}}(\Omega)=\frac{1}{4}\sum_{\bf
k}\left (\gamma^{B_{1g}}_{\bf k}\right)^2 \left \{\frac{\Delta_{AF}^2}{E^2_{\bf
    k}}\Lambda_{\bf k} \delta \left (\Omega -
E^+_{\bf k}-E^-_{\bf k}\right)\right. && {} \nonumber \\ \left.{}
+ \sum_{\tau=\pm}\frac{\Delta_{S,{\bf k}}^2}{(E_{\bf k}^\tau)^2}\left (1+\frac{\tau
\xi^-_{\bf k}}{E_{\bf k}}\right)\left( \frac{\tau \xi^-_{\bf k}}{E_{\bf
  k}}\right ) \delta
   \left (\Omega-2E^\tau_{\bf k}\right ) \right \}&&{},
\label{ramancoexb1g}
\end{eqnarray}
with  $\Lambda_{\bf k}=\left ( 1 - \frac{(\xi_{\bf k}^+)^2+\Delta^2_{S,\bf
      k}-E^2_{\bf k}}{E^+_{\bf k}E^-_{\bf k}}\right )$,
 $\xi^\pm_{\bf
  k}=(1/2)(\xi_{\bf k}\pm \xi_{\bf k+Q})$,
$E_{\bf k}^{\pm}=[(\xi^+_{\bf k}\pm E_{\bf k})^2 +
  (\Delta_{S,\bf k})^2 ]^{1/2}$,
$E_{\bf
  k}=[(\xi^-_{\bf k})^2 + \Delta^2_{AF}]^{1/2}$.
In the calculations we use Eq.~(\ref{formularaman}) with a constant
scattering rate $\Gamma$ except otherwise stated. A more proper
treatment of the scattering rate should include the effects of
impurities, inelastic scattering, disorder and the well-known flat
background at high frequencies but it is beyond the scope of this
article\cite{devereauxprl95}.

{\it Hole-doped cuprates.} We use the Yang-Rice-Zhang
model\cite{YRZ} to describe the pseudogap in hole-doped cuprates.
This model assumes that the pseudogap state can be described as
doped
  resonant valence bond (RVB) state and
proposes an ansatz for the coherent part of the single particle
Green's function to characterize it. This ansatz is proposed in
analogy with the
  form derived for a doped spin liquid formed by an array of two-leg Hubbard
  ladders near half-filling. The description starts from t-J model and uses
  the Gutzwiller
  approximation to project out double occupied sites as in the early
  renormalized mean field description of the RVB state\cite{RMF}.
  In particular the
  kinetic energy (i.e. the values of the hopping parameters)
  and the coherent quasiparticle spectral weight are
  renormalized via
$g_t=2x/(1+x)$ and  depend on doping. We use the doping dependence
of the hopping parameters proposed in\cite{YRZ}. A new feature of
the YRZ model is to describe the pseudogap correlations at zero
temperature by a parameter $\Delta_R$ via a self energy
$\Sigma_R({\bf k},\omega)=\Delta_{R,{\bf k}}^2/(\omega+\xi_{0{\bf
k}})$ which diverges at zero frequency at the umklapp surface
$\xi_{0{\bf k}}$ (Luttinger surface). Here $\xi_{0{\bf
k}}=-2t_0(x)(\cos k_x+\cos k_y)$ and $\Delta_{R,{\bf
k}}=(\Delta_R(x)/2)(\cos k_x -\cos k_y)$. Hole pockets appear close
to $(\pm \pi/2,\pm \pi/2)$ but $\Delta_R$  does not break any
symmetry. In the superconducting state of underdoped cuprates
coexistence of pseudogap and superconductivity is assumed. The
diagonal element of the matrix Green's function becomes
\begin{equation}
G^{RVB}_{SC}({\bf k},\omega)=\frac{g_t}{\omega-\xi({\bf
k})-\Sigma_R({\bf k},\omega)-\Sigma_S({\bf k},\omega)}
\label{eq:greensc}.
\end{equation}
with $\xi_{\bf k}=\xi_{0{\bf k}}-4t_1(x)\cos k_x \cos
k_y-2t_2(x)(\cos 2k_x+\cos 2k_y)-\mu_p$ and $\mu_p$  determined from
the Luttinger sum rule. $\Sigma_S({\bf k},\omega)=|\Delta^2_{S,{\bf
k}}|/(\omega+\xi({\bf k})+\Sigma_R({\bf k},-\omega)$) is the
superconducting self energy with
$\Delta_{S,\bf{k}}=(\Delta_{S}(x)/2)(\cos k_x-\cos k_y)$  the
superconducting order parameter. In the pseudogap state there are
two quasiparticle bands with strongly varying spectral weight. These
two bands become four in the superconducting state, $\pm
E^{\pm,h}_{\bf k}$ \cite{YRZ,nosotras07}:
\begin{eqnarray}
(E^{\pm,h}_{\bf k})^2&=&\Delta_{R{\bf k}}^2+\frac{\xi^2_{\bf
k}+\xi^2_{0{\bf k}}+\Delta_{S{\bf k}}^2}{2}\pm (E_{\bf k
}^{SC,h})^2\nonumber\\
(E_{\bf k}^{SC,h})^2&=&\sqrt{(\xi^2_{\bf k}-\xi^2_{0{\bf
k}}+\Delta_{S{\bf k}}^2)^2+4\Delta_{R{\bf k}}^2((\xi_{\bf
k}-\xi_{0{\bf k}})^2+ \Delta_{S{\bf k}}^2)}.\nonumber
\label{eq:bandshd}
\end{eqnarray}
The Raman spectrum
is calculated in the bubble approximation as in\cite{nosotras07}.

{\it Parameters}.
\begin{figure}
\leavevmode
\includegraphics[clip,width=0.45\textwidth]{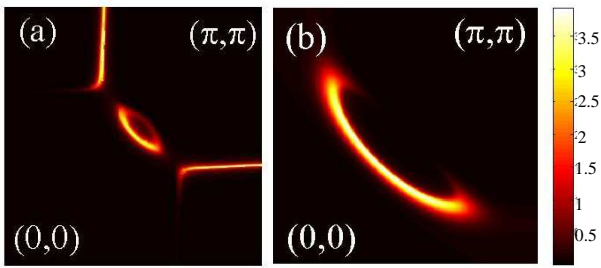}
\caption{(Color online)
Simulated map of ARPES intensity  at the Fermi energy with
  $\Gamma=0.02$ for electron-doped cuprates in (a) and hole-doped cuprates in
  (b) for the parameters given in the text.
The spectral function  is convoluted with a gaussian of width $0.06$
to mimic finite resolution. }
 \label{1}
\end{figure}
In both electron and hole doped cuprates the PG scales $\Delta_{AF}$
and $\Delta_R$ decrease with doping and vanish at a Quantum Critical
Point (QCP) at which Fermi liquid and BCS description are recovered
in the normal and superconducting states, respectively. In
hole-doped compounds the hopping parameters are renormalized
according to the Gutzwiller approximation and depend also on
doping\cite{YRZ}. In order to compare the behavior in electron and
hole-doped systems we keep the doping constant and equal to $x=0.13$
for electron doped and $x=0.14$ in the hole-doped case and vary
$\Delta_S$. We work in units of the unrenormalized  nearest neighbor
hopping $t=1$. For electron-doped cuprates we use $t_0=1$,
$t_1=-0.3$, $t_2=0.25$, $\Delta_{AF}=0.3$ and $\mu=0.59$,
corresponding to $x=0.13$, which reproduces well the ARPES intensity
at the Fermi level (Fig.~1(a)). In hole-doped cuprates we use
$t_0=0.37$, $t_1=-0.073$, $t_2=0.05$, $\Delta_R=0.18$ and
$\mu_p=-0.29$. The ARPES intensity corresponding to these values is
shown in Fig.~1(b). The parameters chosen reproduce reasonably well
the ARPES spectrum for both electron and hole-doped cuprates.
However they are not expected to give a quantitative fitting of the
Raman spectrum. In the hole-doped case the parameters chosen are the
ones originally proposed in the paper by Yang-Rice and
Zhang\cite{YRZ}, and later used in \cite{nosotras07,nosotras08}. To
better compare with the hole-doped case and for numerical
convenience the superconducting order parameter in the
electron-doped case is larger than the experimental one. Our
emphasis is in the qualitatively different behavior observed in the
Raman spectrum of electron and hole-doped cuprates.

The aim of this paper is to compare the qualitative changes of the
Raman spectrum when going from the normal-pseudogap to the
superconducting state. Experimentally the normal state is reached
with increasing temperature above $T_c$. On the other hand, the YRZ
model used for hole-doped cuprates, based on the Gutzwiller
projection, has been developed only for zero temperature. To mimic
the effect of going from the superconducting to the normal state, we
vary $\Delta_S$, keeping the temperature equal to zero and the PG
scale, $\Delta_{AF}$ or $\Delta_R$ constant. We believe that the
  qualitative features described in this paper will be present when going from
the superconducting to the normal state by increasing the
temperature.

\section{Raman Spectrum}

{\it Electron-doped cuprates.}
\begin{figure}
\leavevmode
\includegraphics[clip,width=0.45\textwidth]{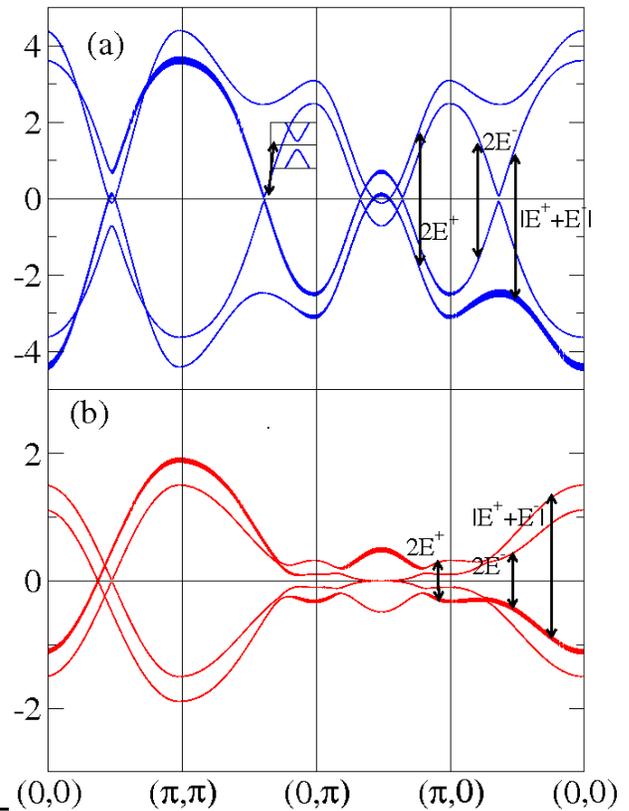}
\caption{(Color online) Energy bands in the superconducting state
corresponding to the
  parameters given in the text for (a)
  electron-doped cuprates and (b) hole-doped cuprates.
The inset in (a) zooms the opening of the
  superconducting gap at the electron pocket along $(0,\pi)$-$(\pi,\pi)$.
The width of the bands gives an
  idea of the corresponding spectral weight. The arrows show the possible
  interband transitions.
}
 \label{2}
\end{figure}
With SC and SDW the energy spectrum consists of four bands with
energies $\pm |E^{\pm}_{\bf k}|$, see Fig.~2(a). The pair breaking
excitations due to superconductivity are given by the terms with
$\delta(\Omega-2E^\pm_{\bf k})$  in (\ref{ramancoexb2g}) and
(\ref{ramancoexb1g}). These terms describe two different
transitions, shown on the rightside of Fig.~2(a), with energies
$\Omega=2E^{\pm}_{\bf k}$. Sketched in the figure  there is a third
kind of interband transition with $\Omega=E^+_{\bf k}+E^-_{\bf k}$.
Its contribution to the Raman spectrum is given by the first term in
Eq.~(\ref{ramancoexb1g}).  In contrast to pair breaking excitations
it is not SC-induced, but it has an SDW origin. A similar transition
shows up also in the Raman response of a SDW in the absence of SC
($\Delta_{S}=0$). Thus this transition remains above $T_c$ if the
temperature at which the AF order appears is larger than $T_c$. As
pointed out in\cite{chinosdelasnarices} the SDW-induced peak is only
seen in $B_{1g}$  and not in $B_{2g}$, both with and without
superconductivity. The corresponding term is missing in
(\ref{ramancoexb2g}). This a consequence of the breaking of symmetry
produced by the SDW and how this symmetry breaking relates to the
polarization of $B_{1g}$ and $B_{2g}$ channels. The different
behavior in both channels arises from the sign which precedes the
$\Pi_{13,31}-\Pi_{14,41}$ term in Eq.~(\ref{formularaman}). Its
contribution to the SDW-induced transition equals, in absolute
value, the one of $\Pi_{11,11}-\Pi_{12,21}$. For $B_{1g}$
polarization both contributions  add while for $B_{2g}$ they cancel
each other. This result also applies to other charge or spin density
wave phases with ${\bf Q}=(\pi,\pi)$\cite{zeyherygreco02}.

The total Raman spectrum is plotted  in the left insets in Fig.~3,
((a) and (c)), for different values of $\Delta_S$. A peak at about
$2\Delta_S$ due to pair breaking excitations is observed in both
$B_{1g}$ and $B_{2g}$ channels. Its intensity decreases and its
position shifts to lower energies as $\Delta_S$ decreases, in a way
which resembles what happens in a BCS superconductor with
$\Delta_{AF}=0$. The low energy feature in the $\Delta_S=0$ curve in
both electron and hole doped cuprates corresponds to the Drude peak
Compared to the BCS spectrum, at low energies the presence of the
SDW suppresses slightly the intensity in the $B_{1g}$ channel since
some of the spectral weight is reorganized due to the opening of the
SDW. The missing spectral weight is found in the SDW-induced peak at
higher energies. In $B_{2g}$ the linear $\Omega$ dependence at low
energies characteristic of a SC gap with nodes is observed. Closer
inspection (see  black curves in the right panels of Fig.~4) shows
that the pair-breaking feature in $B_{2g}$ has a double-peak
structure.  The double peak structure originates in the two
transitions with energies $|2E^+_{\bf  k}|$ and $|2E^-_{\bf k}|$
discussed above. The double peak appears in both $B_{1g}$ and
$B_{2g}$ Raman response though in $B_{1g}$ one of the peaks is very
much suppressed  and cannot be resolved at the scales shown in the
figures. Both peaks are very close in frequency and the scattering
rate makes them to merge into a single one  as can be appreciated in
the $B_{2g}$ response (black dashed curve) of Fig.~4a. The peak at
$\Omega=2\Delta_{AF}$ in the $B_{1g}$ spectrum arises from the
SDW-induced transition.

The effect of superconductivity in the Raman spectrum is better seen
in the main figures in Fig.~3(a) and (c), where the response in the
SDW-normal state has been subtracted. Except for the suppression of
intensity in the $B_{1g}$ spectrum at $\Omega=2 \Delta_{AF}$ no
feature shows up in the subtracted figures at the SDW energy scale.
This dip in intensity is due to the $\left (\frac{\tau \xi^-_{\bf
k}}{E_{\bf k}}\right)$ factor in (\ref{ramancoexb1g}) which makes
that the contribution of the hot spots to the pair breaking $B_{1g}$
intensity vanishes. In the difference spectra the peaks in $B_{1g}$
and $B_{2g}$ appear at the same energy scale and have a similar
dependence on $\Delta_S$. Below we show that this is not the case in
hole-doped cuprates.

\begin{widetext}

\begin{figure}[h!]
\leavevmode
\includegraphics[clip,width=1.0\textwidth]{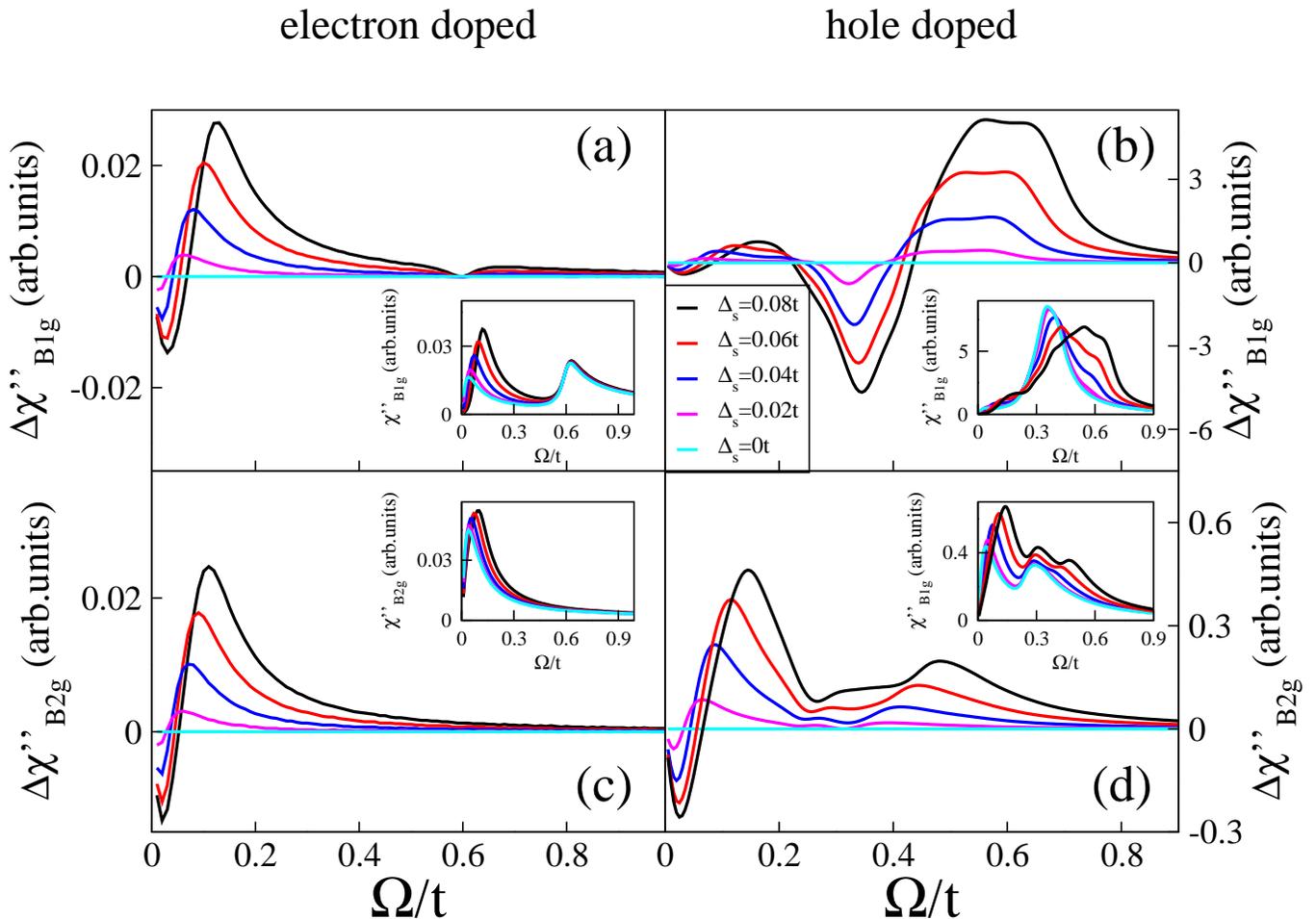}
\caption{ (Color online) Raman response in $B_{1g}$ (top) and $B_{2g}$ (bottom)
channels in electron (left) and hole (right) doped cuprates as a
function of the superconducting order parameter $\Delta_S$. Insets
show the total response $\chi''_{B_{1g},B_{2g}}$ and main figures
the difference one
$\Delta\chi''_{B_{1g},B_{2g}}=\chi''_{B_{1g},B_{2g}}(\Delta_S)-\chi''_{B_{1g},B_{2g}}(\Delta_S=0)$.
Intensity is in arbitrary units. $\Gamma$ is given in units of the
  unrenormalized nearest neighbor hopping parameter}
 \label{3}
\end{figure}

\end{widetext}

{\it Hole-doped cuprates} As to the case of electron-doped cuprates
discussed above the energy spectrum is composed by four bands with
energies $\pm E^{\pm,h}_{\bf k}$, see Fig.~2(b), which expressions
are given by Eq.~\ref{eq:bandshd}. The Raman spectrum is composed by
three transitions: two pair-breaking peaks with energies
$2|E^{-,h}_{\bf k}|$ and $2|E^{+,h}_{\bf k}|$ and a third {\it
crossing} transition with energy $|E^{-,h}_{\bf k}+ E^{+,h}_{\bf
k}|$.  The {\it crossing} transition is associated to the PG
correlations in a similar way as the SDW-transition in the
electron-doped materials is associated to antiferromagnetism.
However, in the hole-doped case,  $\Delta_R$ does not break any
symmetry, $\Pi_{13}$ and $\Pi_{14}$ vanish and the {\it crossing}
transition is active in both $B_{1g}$ and $B_{2g}$.

The Raman spectrum shown in the right insets of Fig.~3, ((b) and
(d)), differs considerably with respect to the one of electron-doped
cuprates. Two peaks appear in $B_{2g}$ channel. The low frequency
one shifts to lower frequencies and decreases in intensity with
decreasing $\Delta_S$ as expected for a pair breaking peak. The
intensity of the weaker high-frequency peak also decreases with
decreasing $\Delta_S$. At $\Delta_S=0$ its intensity is finite and
comes from the {\it crossing} transition discussed above. Note that
as in the electron-doped case the low energy peak which appears
  for $\Delta_S=0$ is due to a finite value of $\Gamma$.

The increase of the high-frequency peak intensity with increasing
$\Delta_S$ indicates that pair breaking excitations also contribute
to the intensity at this frequency in the superconducting state. The
$B_{1g}$ response is dominated by the high-frequency peak whose
intensity is reorganized in the superconducting state. The
low-frequency feature appears just as a small shoulder in the total
spectrum. Interestingly, as experimentally
observed\cite{sacuto08b1g} all the curves seem to cross at a single
point (isosbetic point).

Main figures in Fig.~3(b) and (d) show the difference response
$\Delta\chi''_{B_{1g},B_{2g}}=\chi''_{B_{1g},B_{2g}}(\Delta_S)-\chi_{B_{1g},B_{2g}}(\Delta_S=0)$.
The $B_{2g}$ spectrum is dominated by the low frequency peak.
However a smaller peak at high frequency can be also discerned. The
low and high frequency peaks in the difference curves are both pair
breaking and correspond respectively to the $|2E^-_{\bf k}|$ and
$|2E^+_{\bf k}|$  transitions. The characteristic energy scales of
the peaks differ strongly from the ones found in the electron-doped
case in which both pair-breaking peaks appeared at a scale given by
$\Delta_S$ and could barely be distinguished. The features in the
$B_{1g}$ difference spectrum differ from those found in
  $B_{2g}$. The low
frequency feature is very weak while the high-frequency pair
breaking peak dominates the response. The last one is wide and its
position does not change  much with $\Delta_S$. Note, that
because the position in frequency  at which the {\it crossing} and
high-frequency pair breaking transitions peak are so close,
intensity at this energy is also expected in the PG state in absence
of superconductivity.

\section{Non-monotonic superconducting gap in electron-doped cuprates}

\begin{widetext}

\begin{figure}
\leavevmode
\includegraphics[clip,width=0.8\textwidth]{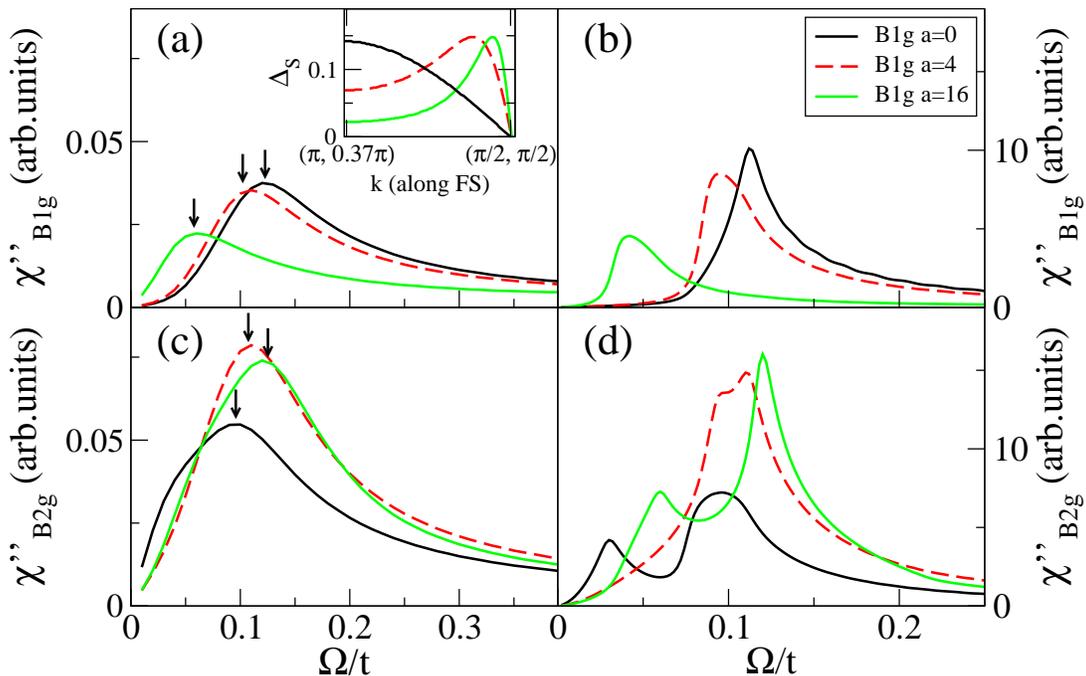}
\caption{ (Color online) (a) and (c) $B_{1g}$ and $B_{2g}$ Raman response
calculated with $\Gamma=0.02$ and corresponding to different
anisotropies of the order parameter given in Eq.~(6) which {\bf
k}-dependence along the Fermi surface (in the absence of SDW gap) is
given in the inset. (b) and (d) Pair breaking peaks corresponding to
the same values as in left figures. This figure has been calculated
using eqs. (3) and (4) and introducing a brodening $0.001$ to the
$\delta$-function. In all the figures a=0 (black solid), a=4 (red
dashed) and a=16 (light green solid). $\Gamma$ and the broadening of
the $\delta$-function are
  measured in units of the unrenormalized nearest neighbor hopping
  parameter. The spectrum is given in
arbitrary units. }
 \label{4}
\end{figure}

\end{widetext}

As discussed in previous section, at the bubble level even in the
presence of a SDW, using a d-wave form for the superconducting order
parameter, the $B_{2g}$ pair-breaking contribution peaks at  a
frequency smaller than the $B_{1g}$ one, contrary to what was
observed at optimal doping in electron-doped
cuprates\cite{blumberg02,qazilbash05}. A non-monotonic d-wave SC gap
is a plausible explanation for the observed peak position. The
spectrum for a non-monotonic SC d-wave gap, in the absence of
antiferromagnetism was reported in\cite{eremin08}. In Fig.~4 we plot
the spectrum for coexisting SDW and a non-monotonic SC order
parameter\cite{eremin08}.
\begin{equation}
\Delta_{S,{\bf k}}=\Delta_0  \frac{\sqrt{a}}{3\sqrt{3}} \frac{\cos k_x-\cos
k_y}{(1+a/4(\cos k_x-\cos k_y)^2)^{3/2}},
\end{equation} for $a=4,16$ which {\bf k}-dependence is given in the inset and
compare it with the spectrum for pure d-wave $\Delta_S$. All the
Raman spectra shown in this figure correspond to the same maximum
value $\Delta_{max}$ but its position shifts towards the node with
increasing $a$. Left panels show the spectrum calculated with
$\Gamma=0.02$, value for which the two pair-breaking peaks cannot be
resolved (as expected experimentally). With increasing $a$ the
position of the $B_{2g}$ peak shifts to larger frequency while the
peak intensity increases. On the other hand the position of the
$B_{1g}$ peak shifts to lower frequency and the peak intensity
decreases. This behavior is expected from the generic form of the
gap and differs  little from the one found for a non-monotonic
d-wave gap in the absence of a SDW\cite{eremin08}. These features
are in agreement with the experimental results. The effect of the
SDW is better seen in the right panels which have been calculated
using Eq.~(3) and (4) adding a
  small broadening to the $\delta$-function.
With increasing $a$ the lowest (highest) pair breaking peak shifts
to higher (lower) frequency. As a consequence, with increasing
non-monotonicity (increasing $a$) the peaks first merge together
($a=4$) and then exchange positions ($a=16$). Due to the similarity
of the spectrum with and without SDW it is not possible to extract
 any fingerprint of the SDW from the low energy Raman spectrum.

\section{Summary and discussion}

  We have analyzed how pseudogap and superconducting scales
  show up in the Raman spectrum of electron and hole-doped
  cuprates. The pseudogap in electron-doped compounds has been described by a
  spin density wave model, while for hole-doped cuprates we have used the
  Yang, Rice and Zhang model for a doped spin liquid.
  In both cases the pseudogap is assumed to remain in the superconducting state
  and to compete with superconductivity removing part of the underlying Fermi
  surface. Below Tc, superconductivity induces pair-breaking peaks in the
  Raman spectrum. Due to the modification of the energy spectrum, in the
  presence of the PG the typical BCS pair-breaking peak splits into two.

  The two pair-breaking peaks are very close in frequency in electron-doped
  systems. One of the peaks is strongly suppressed in the $B_{1g}$ channel.
  The characteristic frequency of both peaks is controlled by the
  superconducting order parameter $\Delta_S$.
  The pseudogap scale $\Delta_{AF}$ does not show up in the
  pair-breaking spectrum.  Except for extremely small (unrealistic) scattering
  rates, the two peaks merge into one and cannot be resolved experimentally
  even in the $B_{2g}$ channel.  An SDW-induced transition is present both
  below and above $T_C$ and produces a peak at $2\Delta_{AF}$. This transition
  is not active in $B_{2g}$ channel.

  In hole-doped compounds the two pair breaking peaks are clearly separated.
The high-energy peak
  frequency is controlled by the maximum gap measured in ARPES in the
  antinodal region\cite{nosotras07}.
  It  depends on both $\Delta_S$ and  the pseudogap scale
  $\Delta_R$. The low-energy peak appears at a frequency
 slightly lower than expected for a BCS superconductor with order
parameter
  $\Delta_S$. It is strongest in $B_{2g}$ but it is barely visible in
  $B_{1g}$.
  A PG induced {\it crossing} transition is present for zero or finite
  $\Delta_S$
  and produces a peak in the spectrum at a frequency which can be
  very close to the
  high-frequency pair breaking peak, making it difficult to disentangle both
  contributions in the total spectrum. This transition has higher intensity
  in $B_{1g}$.  When the superconducting gap decreases, if the pseudogap scale
  does not change,  the high energy peak in $B_{1g}$
  barely changes its position. Similar behavior has been
  found in recent experiments
  by Sacuto's group\cite{sacuto08b1g} for a given sample
  with increasing temperature and for samples with the same doping but
  different critical temperatures due to impurity
  substitution\cite{sacutoimpurezas}.

A convenient way to analyze the effect of superconductivity in the
spectrum is to look at the difference response
$\Delta\chi''_{B_{1g},B_{2g}}=\chi''_{B_{1g},B_{2g}}(\Delta_S)-\chi_{B_{1g},B_{2g}}
  (\Delta_S=0)$.
For realistic scattering rate values, in electron doped cuprates the
effect of PG in $\Delta\chi''_{B_{1g},B_{2g}}$ almost vanishes in
the difference response except for a dip at $2\Delta_{AF}$ in
$B_{1g}$ and a slight change of shape of the pair breaking peaks.
With decreasing $\Delta_S$ the peaks in $\Delta\chi''_{B_{1g}}$ and
$\Delta\chi''_{B_{2g}}$ shift to lower frequencies and decrease
their intensities.  However, two clearly differentiated peaks appear
in the difference response in hole-doped compounds.
$\Delta\chi''_{B_{2g}}$ is dominated by the low frequency one and
behaves qualitatively in the standard way (shift to lower
frequencies and decrease in intensity with decreasing $\Delta_S$).
On the contrary the $\Delta\chi''_{B_{1g}}$ is controlled by the
high-frequency peak. Its intensity decreases with decreasing
$\Delta_S$ but its position barely depends on it (this could change
to some extent if the pseudogap energy scale is very small).

The appearance or not of the PG-induced (SDW or {\it crossing})
transition  in $B_{2g}$ channel in electron and hole-doped
cuprates originates in the
different model used in both cases which folds the Brillouin zone at
the AFZB in electron-doped but does not break any symmetry in the
hole-doped case.

We are not aware of any signature of the SDW transition in the Raman
spectrum of electron doped cuprates. If the SDW model discussed here
is applicable, such a transition would be
  active in the optical conductivity too.
In fact a peak at $\sim 200 meV$ in the optical conductivity in the
normal state has been previously associated to an
SDW\cite{bontemps}. In hole-doped cuprates a broad feature in
$B_{1g}$ above $Tc$ at the same frequency at which the pair-breaking
peak shows up in the superconducting state has been observed and is
considered to be the Raman signature of the
pseudogap\cite{sacuto08b1g}.

On the contrary the different behavior of the energy scales is not a
consequence of the different model used but on the different
truncation of the Fermi surface produced by the pseudogap. Two
energy scales like the ones  discussed here and in\cite{nosotras07}
for hole-doped compounds, would appear in other competing models if
the parameters chosen result in a single nodal Fermi pocket of size
and shape similar to the ones in\cite{nosotras07}. In hole-doped
cuprates the effect of the pseudogap in the spectrum is strongest in
the region of {\bf k}-space which controls the high-frequency peak,
destroying the Fermi surface at the Brillouin zone edge. As a
consequence, the energy of the associated pair-breaking transition
increases but its intensity decreases, because part of the spectral
weight goes to the {\it crossing} transition. As this region, close
to $(\pi,0)$ is mainly sampled by $B_{1g}$ channel, the spectrum in
this channel is highly anomalous. On the contrary $B_{2g}$ mostly
samples the Fermi pocket; in the inner edge, the arc, the spectrum
is more conventional. In electron-doped systems the pseudogap gaps
the Fermi surface at the hot spots, far from $(\pi,0)$. The low and
high-frequency pair breaking peaks originate respectively in  the
hole and electron pockets at nodal and antinodal regions. As there
is a well defined Fermi surface in each pocket, and a gap equal to
$\Delta_{S,{\bf k}}$ opens in each of these Fermi surfaces both
pair-breaking peaks show up at a frequency controlled by $\Delta_S$.

We have also calculated the spectrum for electron-doped compounds
using a non-monotonic d-wave superconducting order parameter in the
presence of a SDW. With increasing non-monotonicity of the
superconducting order parameter the two pair-breaking peaks first
merge together and then exchange positions. Thus, for large
non-monotonicity the spectrum peaks at frequency larger in $B_{2g}$
than in $B_{1g}$.  For realistic values of the scattering rate
the spectrum at low energy is very similar to the one without spin
density wave.

 In conclusion, though competing scenarios might be valid for
hole and electron doped cuprates, its Raman spectrum can be very
different and a careful analysis is needed. While in electron doped
cuprates the superconductivity and  pseudogap are
practically disentangled this is not the case for the hole doped
cuprates. We have shown that the different truncation of the Fermi
surface for hole and electron doped cuprates is key to understand
this completely different behavior. The calculation of the Raman spectrum has
been performed in the bubble approximation. We expect that vertex corrections
would not modify the qualitative behavior discussed here.

Finally we note that recent experiments\cite{park07,motoyama07}
suggest that in electron doped cuprates the antiferromagnetic order
in the superconducting state is short range. The SDW model used here
to characterize electron-doped cuprates assumes long-range
antiferromagnetism. We believe that the results reported here are
still valid for short range interactions. Some features could be
broadened, in a way similar to the non-resolution limited elastic
magnetic Bragg peaks\cite{neutrones}. We also note that for both
electron and hole doped systems we have kept the scattering rate
constant when changing $\Delta_S$. Experimentally the scattering
rate has a non trivial dependence in $\omega$, ${\bf k}$ and $T$
what could influence the spectrum to some extent.

We thank T.M. Rice, A.F. Santander-Syro, T.P. Devereaux and A. Sacuto 
for useful conversations. E.B. thanks the hospitality of the ETH-Zurich and
ESPCI-Paris where part of this work was done. Funding from MCyT
through Grant No. FIS2005-05478-C02-01 and Ramon y Cajal contract
and from Consejeria de Educacion de la CAM and CSIC through Grant
No. CG07-CSIC/ESP-2323 and I3P contract is acknowledged.


\begin{thebibliography}{99}
\bibitem{aclaracion}That the non-monotonicity of the superconducting gap
  cannot be explained as a consequence of the opening of an antiferromagnetic
  gap does not mean that this non-monotonicity does not originate in
  antiferromagnetic fluctuations, if they are the origin of superconductivity,
  as discussed in\cite{krotov06}.
\end{thebibliography}
\end{document}